\documentclass[]{eas}

\usepackage{natbib}
\usepackage{graphicx}

\newcommand{\kms}{\,\rm km\,s^{-1}}

\begin{document}

\title{Magnetic~field~amplification by SN-driven~interstellar~turbulence}
\runningtitle{Gressel \etal: SN-driven dynamo}

\author{Oliver Gressel, Udo Ziegler \& Detlef Elstner}
\address{Astrophysikalisches Institut Potsdam, 
  An der Sternwarte 16, 14482 Potsdam, Germany}

\begin{abstract}{%
  Within the interstellar medium, supernovae are thought to be the prevailing
  agents in driving turbulence. Until recently, their effects on magnetic
  field amplification in disk galaxies remained uncertain. Analytical models
  based on the uncorrelated-ensemble approach predicted that any created field
  would be expelled from the disk before it could be amplified
  significantly. By means of direct simulations of supernova-driven
  turbulence, we demonstrate that this is not the case. Accounting for
  galactic differential rotation and vertical stratification, we find an
  exponential amplification of the mean field on timescales of several hundred
  million years. We especially highlight the importance of rotation in the
  generation of helicity by showing that a similar mechanism based on
  Cartesian shear does not lead to a sustained amplification of the mean
  magnetic field.}
\end{abstract}

\maketitle

\section{Introduction} 

In the framework of the so-called turbulent $\alpha$ effect, rotation has
always been considered the pivot point in the generation of ``cyclonic
turbulence''. Recently, by means of shearing box simulations with peculiarly
elongated aspect ratios, \citet{2008PhRvL.100r4501Y} have claimed that a
turbulent dynamo can already be excited in the presence of shear alone, i.e.,
in the absence of rotation. This finding notably disagrees with quasi-linear
theory \citep{2006AN....327..298R} -- at least for order of unity magnetic
Prandtl numbers.

With this controversy in mind, we want to turn to the question whether the
galactic dynamo indeed depends on rotation as a source of helicity.
\citet{2006AN....327..298R} derive a non-vanishing shear $\alpha$ effect for
the case where there exist gradients in the turbulence. This is certainly the
case for the stratified galactic disk.

\begin{figure} 
  \begin{minipage}[t]{0.69\columnwidth}
    \includegraphics[width=0.85\columnwidth]{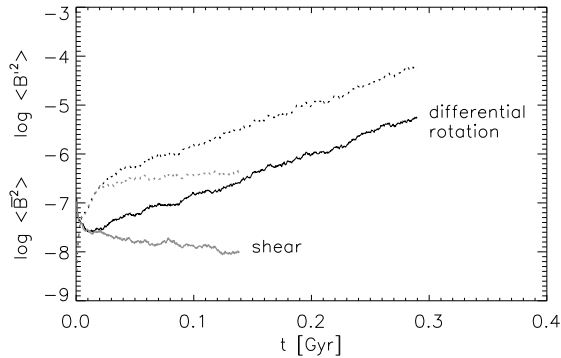}
  \end{minipage}
  \begin{minipage}[b]{0.30\columnwidth}
    \caption{Evolution of the regular (solid) and fluctuating (dashed line)
      magnetic field strength. We compare the cases of differential rotation
      (dark) and shear (light colour).\vspace{5ex}}
    \label{fig:fig1}
  \end{minipage}
\end{figure}

\section{Simulations results} 

In Fig.~\ref{fig:fig1} we see that the irregular field is indeed amplified by
the combined action of turbulence and shear. This means that already the
small-scale dynamo benefits from the local field-line stretching induced by
the shear gradient. The growth, however, happens at a much lower rate compared
to the case of differential rotation. Moreover, under the effect of the
Coriolis force, the mean field grows at a similar rate compared to the
irregular component. In the contrary, we observe a decaying mean magnetic
field in the case of shear alone. As we will see from a detailed analysis of
the inferred dynamo coefficients, this is not because there is no
$\alpha$~effect but because the shear-induced effect has the wrong sign -- for
which the turbulent pumping is too weak to support the dynamo against the
strong galactic wind. Note that the only difference between the models lies in
the Coriolis force; because curvature terms are neglected in the shearing box
approximation, both runs assume the same linear profile of the background
velocity.

\section{Mean-field modelling} 

Our simulations are based on first principles and their outcome has to be
regarded as rather fundamental. Since, however, the current setup only
represents a narrow region in parameter space, our findings have to be backed
up by the underlying theory. Consequently, by means of the so-called
test-field method \citep{2005AN....326..245S}, we derive closure parameters in
the framework of mean-field MHD -- for details see
\citet{2008AN....329..619G}, and references therein.

\begin{figure} 
  \includegraphics[width=\columnwidth,%
    bb=11 104 330 195,clip]{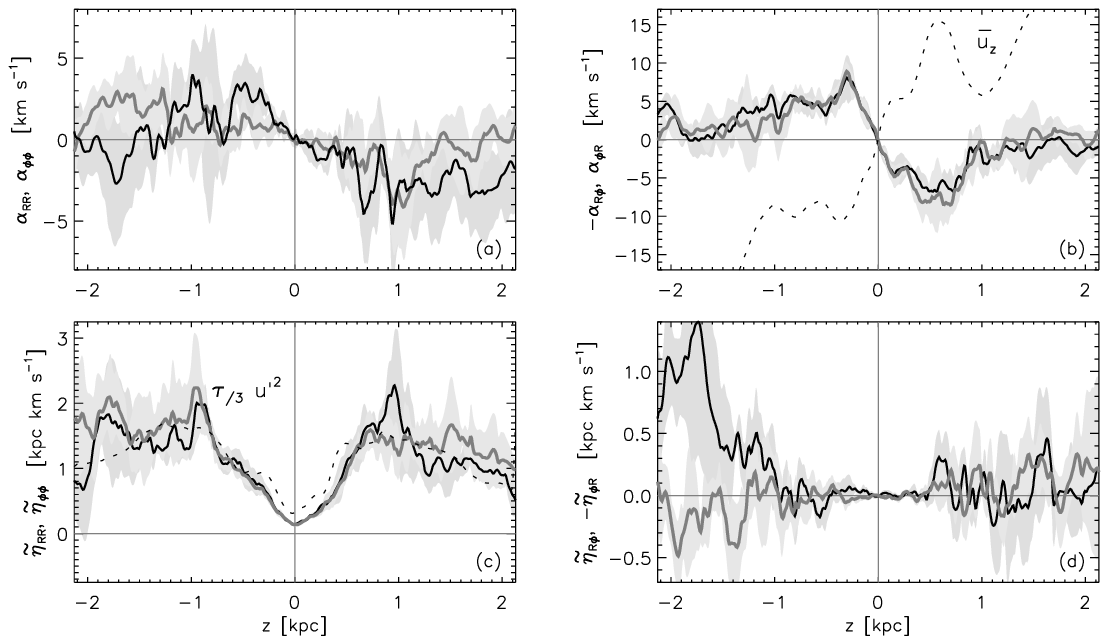}\vskip-4pt%
  \includegraphics[width=\columnwidth,%
    bb=11 009 330 016,clip]{fig/fig2}\vskip-8pt%
  \caption{Components of the $\alpha$ tensor for the case of Cartesian
    shear. The quantities are plotted in dark ($\alpha_{RR},-\alpha_{R\phi}$)
    and light ($\alpha_{\phi\phi}, \alpha_{\phi R}$) colours, respectively.}
  \vspace{8pt}
  \label{fig:fig2}
\end{figure}

The relevant coefficients of the $\alpha$~tensor in the case of Cartesian
shear are presented in Fig.~\ref{fig:fig2}. The diagonal elements (left
panel) clearly show a negative (positive) sign in the top (bottom) half of the
simulation box. While this is opposite to the case of differential rotation
\citep[see Fig.~1 in][]{2008A&A...486L..35G}, the vertical transport via the
mean flow $\bar{u}_z$ and the off-diagonal elements (right panel) are
comparable in both cases.

Based on the inferred $\alpha$ and $\tilde{\eta}$ coefficients, we study the
dependence on the net vertical transport with the help of a 1D dynamo
model. Because the shear gradient defines a distinct sense of orientation, the
overall sign in the $\alpha$~effect is indeed significant. It is well known
from solar dynamo models that the sign of $\alpha\Omega$ determines the
direction of the travelling dynamo wave. In the absence of net vertical
transport, this is reflected in dynamo patterns travelling towards the
midplane for differential rotation, and away from it for shear alone. As we
will see shortly, the different behaviour affects the overall growth rate when
the effects of the diamagnetism and mean flow are included.

\begin{figure} 
  \begin{minipage}[t]{0.69\columnwidth}
    \includegraphics[width=0.9\columnwidth]{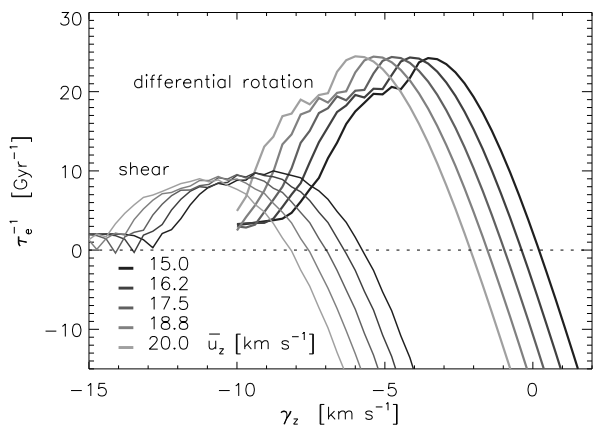}
  \end{minipage}
  \begin{minipage}[b]{0.30\columnwidth}
    \caption{Dynamo growth-rates as a function of the diamagnetic velocity
      $\gamma_z $ for different peak values $\bar{u}_z$ of the galactic
      wind. Negative $\tau_{\rm e}$ correspond to decaying solutions.
      \vspace{5ex}}
    \label{fig:fig3}
  \end{minipage}
\end{figure}

In Fig.~\ref{fig:fig3}, we plot growth rates of dynamo solutions obtained by
varying the amount of the diamagnetic transport $\gamma_z= \frac{1}{2}\,
\left\langle \alpha_{\phi R} - \alpha_{R\phi} \right\rangle$ and the amplitude
of the mean flow $\bar{u}_z$ while keeping the parameters $\alpha_{RR}$,
$\alpha_{\phi\phi}$, and $\eta_{\rm t}$ fixed.  For differential rotation, the
fastest growing solutions are obtained at an outward residual
velocity\footnote{Due to the different shape of the contributing profiles
  (cf. Fig.~\ref{fig:fig2}), the definition of a ``residual'' is not quite
  straightforward. As a rule of thumb, one can double the value for
  $\gamma_z$.} of $\simeq 7\kms$, which agrees well with the findings of
\citet*{1994A&A...286...72S} and \citet{2001A&A...370..635B}. For stronger
inward pumping, the field is more and more squeezed into the midplane and the
effective turbulent dissipation is enhanced. If, on the other hand, we reduce
the amount of pumping, the wind can efficiently remove the created field and
the dynamo mechanism is quenched as well. Independent of $\bar{u}_z$, we find
a limit of $\simeq15\kms$ for the residual velocity, which implies that a weak
inward pumping will already lead to growing dynamo solutions.

In the case of Cartesian shear, the situation is drastically changed. Because
of the opposite sign of the $\alpha$~effect, the dependence on the diamagnetic
pumping is found to be much more critical. Because the basic solution already
constitutes outward travelling dynamo waves, a much stronger inward pumping is
needed to balance the galactic wind. In accordance with the results from our
direct simulations, we consequently do not observe growing dynamo modes
(cf. Fig.~\ref{fig:fig3}) at a realistic level of turbulent pumping.

\section{Conclusions}

By means of combined direct numerical simulations and 1D dynamo models, we
have demonstrated that the $\alpha$~effect arising from the combined action of
stratification and Cartesian shear does not lead to a growing mean magnetic
field in the context of the galactic disk. While the one-dimensional model
closely resembles the elongated simulation box, it has to be checked carefully
to what extent this approach puts constraints on the admissible dynamo
solutions. This means that global mean-field models are required to further
support the current results. With respect to the direct simulations, there
remains the possibility that the relevant effects also depend on the magnetic
Prandtl number ${\rm Pm}$, which needs to be checked via quasi-linear theory.

\end{document}